\documentstyle[preprint,prd,aps]{revtex}

\global\arraycolsep=2pt
\makeatletter
\def\fmslash{\@ifnextchar[{\fmsl@sh}{\fmsl@sh[0mu]}}
\def\fmsl@sh[#1]#2{  \mathchoice
    {\@fmsl@sh\displaystyle{#1}{#2}}    {\@fmsl@sh\textstyle{#1}{#2}}    
{\@fmsl@sh\scriptstyle{#1}{#2}}    {\@fmsl@sh\scriptscriptstyle{#1}{#2}}}
\def\@fmsl@sh#1#2#3{\m@th\ooalign{$\hfil#1\mkern#2/\hfil$\crcr$#1#3$}}
\makeatother

\begin{document}
\draft
\title{{\Large Necessary And Sufficient Condition of Separability of Any System}}
\author{Ping-Xing Chen and Lin-Mei Liang}
\address{Laboratory of Quantum Communication and Quantum Computation, \\
University of Science and Technology of\\
China, Hefei, 230026, P. R. China \\
and \\
\thanks{%
Corresponding address}Department of Applied Physics, National University of\\
Defense Technology,\\
Changsha, 410073, \\
P. R. China}
\author{Cheng-Zu Li}
\address{Department of Applied Physics,\\
National University of Defense Technology,\\
Changsha, 410073, P. R. China }
\author{Ming-Qiu Huang}
\address{CCAST (World Laboratory) P.O. Box 8730, Beijing, 100080, China\\
and Department of Applied Physics,\\
National University of Defense Technology,\\
Changsha, 410073, P. R. China }
\date{\today}
\maketitle

\begin{abstract}
The necessary and sufficient condition of separability of a mixed state of
any systems is presented, which is practical in judging the separability of
a mixed state. This paper also presents a method of finding the disentangled
decomposition of a separable mixed state.
\end{abstract}


\thispagestyle{empty}

\newpage \pagenumbering{arabic} 

Entanglenment is an important element both in quantum theory\cite{1} and in
quantum information\cite{2,3,b11,b12}. A mixed state $\rho $, which acts on
Hilbert space $H_1\otimes H_2$, has no entanglement or is separable iff $%
\rho $ can be written as the following form\cite{4} 
\begin{equation}
\rho =\sum_ip_i\rho _i^1\otimes \rho _i^2  \label{def1}
\end{equation}
where $\rho _i^1$ and $\rho _i^2$ are states of Hilbert space $H_1$ and $H_2$%
, respectively, and $p_i\geq 0$, $\sum p_i=1$. A separable system satisfies
the Bell inequality, but the converse is not necessarily true\cite{5}. Peres
has shown in Ref.\cite{6} that a necessary condition (NC) for separability
of a state $\rho $ is the positivity of its partial transformation $\rho ^T$%
. Here $\rho ^T$ is defined as 
\begin{equation}
\rho _{m\mu ,n\nu }^{T_1}\equiv \rho _{n\mu ,m\upsilon },\qquad \rho _{m\mu
,n\nu }^{T_2}\equiv \rho _{m\nu ,n\mu }.  \label{con1}
\end{equation}
It has been shown that for $2\times 2$ systems (the dimensions of $H_1$ and $%
H_2$ are 2 and 2, respectively) and $2\times 3$ systems, the condition (\ref
{con1}) is also a sufficient one\cite{7}. A necessary and sufficient
condition(NSC) for $m\times n$ systems with the aid of the concept of a map
has been presented by Horodecki {\it et.al} in Ref.\cite{7}, but it is
difficult to judge whether a state is separable or not in practice with this
method. W.D$\stackrel{..}{u}$r {\it et.al }\cite{8} show a NSC of a certain
family of mixed states. In this paper, we shall present a NSC of
separability of $\rho $ for any system. Our main idea is based on the idea
proposed by Wootters in Ref.\cite{9,10}. Compared with Horodecki {\it et.al}%
's work, the NSC presented in this paper is more explicitly algorithmic, so
it is practical in operation. The organization of this paper is as follows.
First, we collect some facts related to our result. Second, the NSC of any
mixed state is presented. Meanwhile, how to get the disentangled
decomposition of a separable state is also presented. Finally, we present an
example of a bound entangled state as an illustration. From this example,
one can know how to judge whether a mixed state is separable or not.

In order to get the results, we collect the facts as follows:

1.Any mixed state $\rho $ of $m\times n$ systems which has $l$ nonzero
eigenvalues can be decomposed into $k$ pure states. And the number of
possible decompositions of $\rho $ is infinite. These decompositions can be
obtained by some transformation $u_{k\times l}$ whose columns are
orthonormal vectors ($k$ is greater than or equal to $l$) \cite{9}. For
example, 
\begin{equation}
\rho =\sum_{i=1}^l\left| x_i\right\rangle \left\langle x_i\right|
=\sum_{i=1}^k\left| z_i\right\rangle \left\langle z_i\right| ,  \label{def2}
\end{equation}
where $\left| x_i\right\rangle ,$ unnormalized, is a complete set of
orthogonal eigenvectors corresponding to the nonzero eigenvalues of $\rho $,
and $\left\langle x_i\right| \left. x_i\right\rangle $ is equal to the ith
nonzero eigenvalue. Then the decomposition $\left| z_i\right\rangle $ of $%
\rho $ can be given by 
\begin{equation}
\left| z_i\right\rangle =\sum_{j=1}^lu_{ij}\left| x_j\right\rangle ,\qquad
i=1,2,\cdots ,k  \label{def2'}
\end{equation}
where the vectors $\left| z_i\right\rangle $ are not necessarily orthogonal.

2. A pure state $\left| \Psi \right\rangle $ of an $m\times n$ systems is
denoted as 
\begin{equation}
\left| \Psi \right\rangle =A_{11}\left| 11\right\rangle +A_{12}\left|
12\right\rangle +\cdots +A_{1n}\left| 1n\right\rangle +\cdots +A_{m1}\left|
m1\right\rangle +\cdots +A_{mn}\left| mn\right\rangle ,  \label{def3}
\end{equation}
where $\left| 11\right\rangle ,\cdots \left| 1n\right\rangle ,\cdots \left|
m1\right\rangle ,\cdots \left| mn\right\rangle $ are the standard basis
vectors of the $m\times n$ system. We introduce the ``vector'' $%
\overrightarrow{A_i}$ 
\begin{equation}
\overrightarrow{A_i}=(A_{i1},A_{i2},\cdots A_{in}).\qquad i=1,2,\cdots ,m
\label{def4}
\end{equation}
It is not difficult to find that the pure state $\left| \Psi \right\rangle $
is separable if and only if all the ``vectors'' $\overrightarrow{A_i}$ are
parallel with another ,which is denoted as 
\begin{equation}
\overrightarrow{A_1}\parallel \overrightarrow{A_2}\parallel \cdots \parallel 
\overrightarrow{A_m}.  \label{deff5}
\end{equation}
Note that the relation(\ref{deff5}) is true if and only if the following is
true 
\begin{equation}
\left\{ 
\begin{array}{c}
(A_{11},A_{12})\parallel (A_{21},A_{22})\parallel \cdots \parallel
(A_{m1},A_{m2}) \\ 
(A_{11},A_{13})\parallel (A_{21},A_{23})\parallel \cdots \parallel
(A_{m1},A_{m3}) \\ 
\vdots \\ 
(A_{11},A_{1n})\parallel (A_{21},A_{2n})\parallel \cdots \parallel
(A_{m1},A_{mn})
\end{array}
\right. .  \label{deff6}
\end{equation}
That is to say there exist $(m-1)(n-1)$ parallel pairs. For example, having
parallel pairs $(A_{11},A_{12})\parallel
(A_{21},A_{22}),(A_{11},A_{12})\parallel (A_{31},A_{32})$ means that $%
A_{11}/A_{21}=A_{12}/A_{22},$ $A_{11}/A_{31}=A_{12}/A_{32}$, respectively.

3. We define $\left| \widetilde{\Psi }^r\right\rangle $ and $c_r$ with
respect to each parallel pair as follows: 
\begin{equation}
\left| \widetilde{\Psi }^r\right\rangle =B^r\left| \Psi ^{*}\right\rangle
,c_r=\left\langle \Psi \right| \left. \widetilde{\Psi }^r\right\rangle
,\quad r=1,2,\cdots ,(m-1)(n-1)  \label{def9}
\end{equation}
where $B^r$ is a $mn\times mn$ symmetric transformation matrix in the
standard basis and there are only 4 nonzero elements of $B^r$. For example,
for $(A_{11},A_{12})\parallel (A_{21},A_{22})$, $\left[ B^1\right]
_{1,n+2}=\left[ B^1\right] _{n+2,1}=-1,\left[ B^1\right] _{2,n+1}=\left[
B^1\right] _{n+1,2}=1$, the other elements of $\left[ B^1\right] $ are all
zero. It is obvious that $\left| c_1\right| =0$ if and only if $%
(A_{11},A_{12})\parallel (A_{21},A_{22}).$ Similarly, each parallel pair in
the relation(\ref{deff6}) can be expressed by its $\left| c_r\right| $ being
equal to zero and this is also the reason we choose the elements of the
matrix $B^r$. For $4\times 4$ systems, there exists nine transformation
matrices as follows:

$(A_{11},A_{12})\parallel (A_{21},A_{22})\rightarrow \left[ B^1\right]
_{1,6}=-1,\left[ B^1\right] _{2,5}=1,\left[ B^1\right] _{5,2}=1,\left[
B^1\right] _{6,1}=-1,$

$(A_{11},A_{12})\parallel (A_{31},A_{32})\rightarrow \left[ B^2\right]
_{1,10}=-1,\left[ B^2\right] _{2,9}=1,\left[ B^2\right] _{9,2}=1,\left[
B^2\right] _{10,1}=-1,$

$(A_{11},A_{12})\parallel (A_{41},A_{42})\rightarrow \left[ B^3\right]
_{1,14}=-1,\left[ B^3\right] _{2,13}=1,\left[ B^3\right] _{13,2}=1,\left[
B^3\right] _{14,1}=-1,$

$(A_{11},A_{13})\parallel (A_{21},A_{23})\rightarrow \left[ B^4\right]
_{1,7}=-1,\left[ B^4\right] _{3,5}=1,\left[ B^4\right] _{5,3}=1,\left[
B^4\right] _{7,1}=-1,$

$(A_{11},A_{13})\parallel (A_{31},A_{33})\rightarrow \left[ B^5\right]
_{1,11}=-1,\left[ B^5\right] _{3,9}=1,\left[ B^5\right] _{9,3}=1,\left[
B^5\right] _{11,1}=-1,$

$(A_{11},A_{13})\parallel (A_{41},A_{43})\rightarrow \left[ B^6\right]
_{1,15}=-1,\left[ B^6\right] _{3,13}=1,\left[ B^6\right] _{13,3}=1,\left[
B^6\right] _{15,1}=-1,$

$(A_{11},A_{14})\parallel (A_{21},A_{24})\rightarrow \left[ B^7\right]
_{1,8}=-1,\left[ B^7\right] _{4,5}=1,\left[ B^7\right] _{5,4}=1,\left[
B^7\right] _{8,1}=-1,$

$(A_{11},A_{14})\parallel (A_{31},A_{34})\rightarrow \left[ B^8\right]
_{1,12}=-1,\left[ B^8\right] _{4,9}=1,\left[ B^8\right] _{9,4}=1,\left[
B^8\right] _{12,1}=-1,$

$(A_{11},A_{14})\parallel (A_{41},A_{44})\rightarrow \left[ B^9\right]
_{1,16}=-1,\left[ B^9\right] _{4,13}=1,\left[ B^9\right] _{13,4}=1,\left[
B^9\right] _{16,1}=-1,$

the other elements of $\left[ B^r\right] $ ($r=1,\cdots ,9$) are all zero.
Similarly, for higher dimension systems, we choose $B^r$ of each parallel
pairs so that each $c_r$ is zero.

4. From the above facts, one can get that any state $\rho $ is separable iff
there is a pure state decomposition of $\rho ,$ $\rho =\sum_i\left|
z_i\right\rangle \left\langle z_i\right| $ with each $\left|
z_i\right\rangle $ being separable, {\it i.e.}, each pair in Eq.(\ref{deff6}%
) is parallel for all $\left| z_i\right\rangle .$

In the following, we will discuss the NSC of separability of $m\times n$
systems. Firstly, we will prove the following lemma.

Lemma. For any density matrix $\rho $ of $m\times n$ systems in Eq.(\ref
{def2}), the statement $(A_{11},A_{12})\parallel (A_{21},A_{22})$ is true
for each pure state $\left| z_i\right\rangle $ in Eq.(\ref{def2'}) if and
only if 
\begin{equation}
a^1=\lambda _1-\sum_{i=2}^{l^{\prime }}\lambda _i\leq 0,  \label{rel1}
\end{equation}
where $l^{\prime }\leq l$ ($l,$ $l^{\prime }$ is the number of eigenvalues
of $\rho ,$ $\tau ^1(\tau ^1)^{*}$ respectively.) and $\lambda _i$s, in
decreasing order, are the square roots of eigenvalues of the matrix $\tau
^1(\tau ^1)^{*}$, where $\tau _{ij}^1=\left\langle x_i\right| \left. 
\widetilde{x}_j^1\right\rangle $ and $\left| x_i\right\rangle $ is defined
in Eq.(\ref{def2}) and $\left| \widetilde{x}_j^1\right\rangle $ in Eq.(\ref
{def9}).

Proof: Obviously $\tau ^1$ is a symmetric matrix. For the symmetric matrix $%
\left[ \tau _{ij}^1\right] ,$ one can find a $l\times l$ unitary matrix $%
\left[ \nu _{ij}\right] $ which can diagonalize the $\left[ \tau
_{ij}^1\right] $ in the following way 
\begin{equation}
\left\langle y_i\right| \left. \widetilde{y}_j^1\right\rangle =\left[ v\tau
^1v^T\right] _{ij}=\lambda _i\delta _{ij},  \label{rel3}
\end{equation}
where $\left| y_i\right\rangle =\sum_j\nu _{ij}^{*}\left| x_j\right\rangle ,$
$\nu _{ij}^{*}$ is the complex conjugation of $\nu _{ij}$, and $\lambda _i$s
are the square roots of eigenvalues of matrix $\tau ^1(\tau ^1)^{*}$\cite{9}%
. Though $\tau ^1$ is dependent on the choice of eigenvectors $\left|
x_i\right\rangle $ of $\rho ,$ $\lambda _i$ is not. Every decomposition of $%
\rho $, $\rho =\sum_i\left| z_i\right\rangle \left\langle z_i\right| ,$ can
be obtained by $\left| z_i\right\rangle =\sum_j\mu _{ij}^{*}\left|
y_j\right\rangle $. Here 
\begin{equation}
\mu ^{*}=\left[ 
\begin{array}{c}
a_{1,1}e^{i\theta _{1,1}}\hspace{0.05in}\hspace{0.31in}\;a_{1,2}e^{i\theta
_{1,2}}\hspace{0.4in}\cdots \qquad a_{1,l}e^{i\theta _{1,l}} \\ 
a_{2,1}e^{i\theta _{2,1}}\hspace{0.05in}\hspace{0.31in}\;a_{2,2}e^{i\theta
_{2,2}}\hspace{0.4in}\cdots \qquad a_{2,l}e^{i\theta _{2,l}} \\ 
\vdots \\ 
a_{k,1}e^{i\theta _{k,1}}\;\quad \;a_{k,2}e^{i\theta _{k,2}}\qquad \cdots
\qquad \;\;a_{k,l}e^{i\theta _{k,l}}
\end{array}
\right]  \label{rel4}
\end{equation}
whose columns are orthonormal vectors and there is no loss of generality in
taking each $a_{ij}$ to be real and nonnegative. If $l^{\prime }<l$, one can
add ($l-l^{\prime }$) dummy states $\left| y_{l^{\prime }+1}\right\rangle
,\left| y_{l^{\prime }+2}\right\rangle ,\cdots ,\left| y_l\right\rangle $
being equal to zero vectors. Thus it can be obtained that $\left|
z_i\right\rangle =\sum_ju_{ij}^1\left| x_j\right\rangle $, where $u^1=\mu
^{*}\nu ^{*}$(there may exist many transformation matrices which can realize
this kind of decomposition. The set which contains all these transformation
matrices is denoted by $U_{k\times l}^1$). If $(A_{11},A_{12})\parallel
(A_{21},A_{22})$ for each $\left| z_i\right\rangle ,$ from fact 3, one has 
\begin{equation}
\left\langle z_i\right| \left. \widetilde{z}_i^1\right\rangle =0,
\label{rel6}
\end{equation}
where $\left| \widetilde{z}_i^1\right\rangle =B^1\left| z_i^{*}\right\rangle 
$; then one has 
\begin{eqnarray}
a_{i1}^4\lambda _1^2 &=&(\sum_{j=2}^{l^{\prime }}a_{ij}^2\cos 2\theta
_{ij}\lambda _j)^2+(\sum_{j=2}^{l^{\prime }}a_{ij}^2\sin 2\theta
_{ij}\lambda _j)^2  \nonumber \\
&\leq &(\sum_{j=2}^{l^{\prime }}a_{ij}^2\lambda _j)^2.  \label{rel7}
\end{eqnarray}
Because of the positivity of $\lambda _i$ and $a_{ij}$, one can get 
\begin{equation}
a_{i1}^2\lambda _1\leq \sum_{j=2}^{l^{\prime }}a_{ij}^2\lambda _j,
\label{rel7'}
\end{equation}
and then, $\lambda _1=\sum_{i=1}^ka_{i1}^2\lambda _1\leq
\sum_{i=1}^k\sum_{j=2}^{l^{\prime }}a_{ij}^2\lambda _j=\sum_{j=2}^{l^{\prime
}}\lambda _j,$ {\it i.e.} 
\begin{equation}
a^1=\lambda _1-\sum_{i=2}^{l^{\prime }}\lambda _i\leq 0.  \label{bbbbb}
\end{equation}

Here we have used Eq.(\ref{rel7'}) and $\sum_{i=1}^ka_{i1}^2=1.$

Conversely, if $\lambda _1\leq \sum_{i=2}^{l^{\prime }}\lambda _i$, one can
find a kind of decomposition of $\rho ,$ $\rho =\sum_i\left|
z_i\right\rangle \left\langle z_i\right| $ [see Appendix], where $%
\left\langle z_i\right| \left. \widetilde{z}_i^1\right\rangle =0,$ so that $%
(A_{11},A_{12})\parallel (A_{21},A_{22})$. Therefore the lemma is proved. 
\begin{tabular}{l}
\end{tabular}

Similarly, from the proof of the Lemma one knows that for an $m\times n$
system the NSC for each of the $(m-1)(n-1)$ pairs being parallel is $a^r\leq
0(r=1,2,\cdots ,(m-1)(n-1))$. If each $a^r\leq 0$, there exist $(m-1)(n-1)$
sets of transformation matrices, $U_{k\times l}^1,U_{k\times l}^2,\cdots
,U_{k\times l}^{(m-1)(n-1)}$, each of which yields decompositions of $\rho $
with the existence of the corresponding parallel pair. Thus we get the main
result in this paper:

Theorem. A NSC of separability of states of an $m\times n$ system is that
the corresponding $a^r\leq 0\quad (r=1,\cdots ,(m-1)(n-1))$ and the
intersection $U_{k\times l}$ of the $(m-1)(n-1)$ sets of transformations is
not empty.

Proof: The sufficient condition is obvious. The task left is to prove the
necessary condition. If $\rho $ is separable , there must exist a
transformation matrix $u$ which can transform the decomposition of $\rho $, $%
\rho =\sum_{i=1}^l\left| x_i\right\rangle \left\langle x_i\right| $ into $%
\rho =\sum_i\left| z_i\right\rangle \left\langle z_i\right| $ with the
existence of $(n-1)(m-1)$ parallel pairs in each pure state $\left|
z_i\right\rangle $. From the Lemma, one can get each $a^r$ $\leq 0$ in each $%
\left| z_i\right\rangle $, and the matrix $u$ belongs to the intersection $%
U_{k\times l}$; {\it i.e. }$U_{k\times l}$ is not empty. Thus the theorem is
proved.

According to the above theorem, one may take the following steps (i)-(iv) to
judge whether a state $\rho $ of an $m\times n$ quantum system is separable
or not.

i). Calculate the nonzero eigenvalues and eigenvectors of $\rho $ (see Eq.(%
\ref{def2})).

ii). Calculate $(m-1)(n-1)$ matrices $\tau ^r(\tau ^r)^{*}$ according to $%
B^r(r=1,2,\cdots ,(m-1)(n-1))$ and obtain the square roots of the
eigenvalues of $\tau ^r(\tau ^r)^{*}$ (see Eq.(\ref{rel3})).

iii). Judge whether $a^r(r=1,2,\cdots ,(m-1)(n-1))$ is greater than zero or
not (see Eq.(\ref{rel1})). If all $a^r\leq 0,$ then go to step (iv),
otherwise $\rho $ is inseparable.

iv). Judge whether the intersection of $(m-1)(n-1)$ sets of transformation $%
U_{k\times l}$ is empty or not. If the intersection is empty, $\rho $ is
inseparable, otherwise $\rho $ is separable.

The steps from (i) to (iii) are easy to follow, but step (iv) is very
difficult to operate at first sight. It is natural to ask whether step (iv)
can be done or not and whether the method is practical . To answer this
question and illustrate the method, we give an example of a bound entangled
state.

A state $\rho $ of a $2\times 4$ system\cite{4} is 
\begin{equation}
\rho =\frac 18\left[ 
\begin{array}{llllllll}
1 & 0 & 0 & 0 & \quad 0 & 1 & 0 & \quad 0 \\ 
0 & 1 & 0 & 0 & \quad 0 & 0 & 1 & \quad 0 \\ 
0 & 0 & 1 & 0 & \quad 0 & 0 & 0 & \quad 1 \\ 
0 & 0 & 0 & 1 & \quad 0 & 0 & 0 & \quad 0 \\ 
0 & 0 & 0 & 0 & 1 & 0 & 0 & 0 \\ 
1 & 0 & 0 & 0 & \quad 0 & 1 & 0 & \quad 0 \\ 
0 & 1 & 0 & 0 & \quad 0 & 0 & 1 & \quad 0 \\ 
0 & 0 & 1 & 0 & 0 & 0 & 0 & 1
\end{array}
\right] .  \label{s16}
\end{equation}
$\rho $ has five nonzero eigenvalues $t_i(i=1,2,\cdots ,5),$%
\begin{eqnarray}
t_1 &=&t_2=\frac 18,  \nonumber \\
t_3 &=&t_4=t_5=\frac 14,  \label{s17}
\end{eqnarray}
The corresponding unnormalized eigenvectors $\left| x_i\right\rangle $ can
be 
\begin{eqnarray}
\left| x_1\right\rangle  &=&(0,0,0,\sqrt{\frac 18},0,0,0,0),\left|
x_2\right\rangle =(0,0,0,0,\sqrt{\frac 18},0,0,0),  \nonumber \\
\left| x_3\right\rangle  &=&(\sqrt{\frac 18},0,0,0,0,\sqrt{\frac 18}%
,0,0),\left| x_4\right\rangle =(0,\sqrt{\frac 18},0,0,0,0,\sqrt{\frac 18},0),
\label{s18'} \\
\left| x_5\right\rangle  &=&(0,0,\sqrt{\frac 18},0,0,0,0,\sqrt{\frac 18}). 
\nonumber
\end{eqnarray}
The three matrices $B^r(r=1,2,3)$ corresponding to the three parallel pairs
are 
\begin{eqnarray}
(A_{11},A_{12}) &\parallel &(A_{21},A_{22})\rightarrow
B_{1,6}^1=B_{6,1}^1=-1,B_{2,5}^1=B_{5,2}^1=1,  \nonumber \\
(A_{11},A_{13}) &\parallel &(A_{21},A_{23})\rightarrow
B_{1,7}^2=B_{7,1}^2=-1,B_{3,5}^2=B_{5,3}^2=1,  \nonumber \\
(A_{11},A_{14}) &\parallel &(A_{21},A_{24})\rightarrow
B_{1,8}^3=B_{8,1}^3=-1,B_{4,5}^3=B_{5,4}^3=1.  \label{S}
\end{eqnarray}
Then the symmetric matrices $\tau _{ij}^r=\left\langle x_i\right| \left. 
\widetilde{x}_j^r\right\rangle =\left\langle x_i\right| B^r\left|
x_j^{*}\right\rangle \quad (i,j=1,2,\cdots ,5,r=1,2,3)$ are 
\begin{equation}
\tau ^1=\left[ 
\begin{array}{lllll}
0 & 0 & \,\,\;0 & 0 & 0 \\ 
0 & 0 & \,\;0 & \frac 18 & 0 \\ 
0 & 0 & -\frac 14 & 0 & 0 \\ 
0 & \frac 18 & \;\,0 & 0 & 0 \\ 
0 & 0 & \,\;0 & 0 & 0
\end{array}
\right] ,\tau ^2=\left[ 
\begin{array}{lllll}
0 & 0 & \;0 & \;0 & 0 \\ 
0 & 0 & \;0 & \;0 & \frac 18 \\ 
0 & 0 & \;0 & -\frac 18 & 0 \\ 
0 & 0 & -\frac 18 & \;0 & 0 \\ 
0 & \frac 18 & \;0 & \;0 & 0
\end{array}
\right] ,\tau ^3=\left[ 
\begin{array}{lllll}
0 & \frac 18 & \;\,0 & 0 & \;0 \\ 
\frac 18 & 0 & \,\;0 & 0 & \;0 \\ 
0 & 0 & \,\;0 & 0 & -\frac 18 \\ 
0 & 0 & \,\;0 & 0 & \;0 \\ 
0 & 0 & -\frac 18 & 0 & \;0
\end{array}
\right] .  \label{s19}
\end{equation}
It is easy to get that $a^1=0,a^2<0,a^3<0,$ so we must turn to step (iv).
According to the theorem, there are surely three sets of transformation
matrices $u_{k\times 5}$ each of which can result in the decomposition $%
\left| z_i\right\rangle $ of $\rho $, $\rho =\sum_{i=1}^k\left|
z_i\right\rangle \left\langle z_i\right| $ with the corresponding pair being
parallel . The $u_{k\times 5}$ can be written as 
\begin{equation}
u_{k\times 5}=\left[ 
\begin{array}{c}
a_{11}\hspace{0.05in}\hspace{0.31in}\;a_{12}\hspace{0.4in}\cdots \qquad
a_{15} \\ 
a_{21}\hspace{0.05in}\hspace{0.31in}\;a_{22}\hspace{0.4in}\cdots \qquad
a_{25} \\ 
\vdots  \\ 
a_{k1}\;\quad \;\quad a_{k2}\qquad \cdots \qquad \quad a_{k5}
\end{array}
\right] .  \label{s20}
\end{equation}
Each column in $u_{k\times 5}$ is an orthonormal vector. If three sets of $%
u_{k\times 5}$ have intersection, the state $\rho $ is separable. This
condition is equivalent to the existence of a set $a_{11},\cdots
,a_{15},\cdots ,a_{k1,\cdots ,}a_{k5}$ satisfying the following equations 
\begin{equation}
\left\{ 
\begin{array}{l}
\left| z_i\right\rangle =a_{i1}\left| x_1\right\rangle +a_{i2}\left|
x_2\right\rangle +\cdots +a_{i5}\left| x_5\right\rangle , \\ 
\left\langle z_i\right| B^1\left| z_i^{*}\right\rangle =0, \\ 
\left\langle z_i\right| B^2\left| z_i^{*}\right\rangle =0, \\ 
\left\langle z_i\right| B^3\left| z_i^{*}\right\rangle =0.
\end{array}
\right. \qquad (i=1,2,\cdots ,k)  \label{s21}
\end{equation}
From Eqs.(\ref{s19}) and (\ref{s21}), one can get that 
\begin{equation}
\left\{ 
\begin{array}{c}
a_{i2}^{*}a_{i4}^{*}=a_{i3}^{*2}, \\ 
a_{i2}^{*}a_{i5}^{*}=a_{i3}^{*}a_{i4}^{*}, \\ 
a_{i1}^{*}a_{i2}^{*}=a_{i3}^{*}a_{i5}^{*}.
\end{array}
\right. \qquad   \label{s22}
\end{equation}
The complex number $a_{ij}^{*}$ can be expressed as $a_{ij}^{*}=b_{ij}e^{i%
\theta _{ij}}$ where $b_{ij}$ is a real and positive number. From the first
equation of Eq.(\ref{s22}), one can get that $\sum_{i=1}^kb_{i2}b_{i4}=%
\sum_{i=1}^kb_{i3}^2=1.$ Because $\sum_{i=1}^kb_{ij}^2=1,$ the equality $%
b_{i2}=b_{i3}=b_{i4}$ can be obtained. Then Eq.(\ref{s22}) results in 
\begin{equation}
\left\{ 
\begin{array}{c}
b_{i1}=b_{i2}=b_{i3}=b_{i4}=b_{i5}=b_i, \\ 
\theta _{i3}=\frac{\theta _{i1}+3\theta _{i2}}4,\theta _{i4}=\frac{\theta
_{i1}+\theta _{i2}}2,\theta _{i5}=\frac{3\theta _{i1}+\theta _{i2}}4.
\end{array}
\right.   \label{s24}
\end{equation}
Because each column in Eq.(\ref{s20}) is orthonormal, one can obtain that 
\begin{equation}
\left\{ 
\begin{array}{c}
\sum_{i=1}^kb_i^2e^{i(\theta _{i1}-\theta _{i2})}=0, \\ 
\sum_{i=1}^kb_i^2e^{i(\theta _{i1}-\theta _{i3})}=\sum_{i=1}^kb_i^2e^{i\frac{%
3(\theta _{i1}-\theta _{i2})}4}=0, \\ 
\sum_{i=1}^kb_i^2e^{i(\theta _{i1}-\theta _{i4})}=\sum_{i=1}^kb_i^2e^{i\frac{%
\theta _{i1}-\theta _{i2}}2}=0.
\end{array}
\right.   \label{s25}
\end{equation}
It is easy to find that Eq.(\ref{s25}) has no solution since $%
\sum_{i=1}^kb_i^2e^{i(\theta _{i1}-\theta _{i2})}=0,\sum_{i=1}^kb_i^2e^{i%
\frac{3(\theta _{i1}-\theta _{i2})}4}=0,$ and $\sum_{i=1}^kb_i^2e^{i\frac{%
\theta _{i1}-\theta _{i2}}2}=0$ do not hold true simultaneously.. Therefore
the intersection $u_{k\times 5}$ of three sets of transformation matrix is
empty and $\rho $ in Eq.(\ref{s16}) is inseparable.

From the above example, one can see that for any mixed state the problem of
finding the intersection of our sets of transformation matrices comes down
to the problem of finding the solution of a set of quadratic equations with
the condition of the solution being orthonormal vectors. The condition of
orthonormality that makes the quadratic equations more easily solved.
Especially, if there are not many nonzero elements of the symmetry matrix $%
\tau ^r$, it is very easy to solve the set of quadratic equations. In
addition, the form of these equations is dependent on the choice of
eigenvectors of $\rho $, which implies that one may simplify the form of
these equations with appropriate eigenvectors of $\rho .$

In summary, we have presented a new necessary and sufficient condition for
separability of state of any system. According to the above theorem, $%
(m-1)\times (n-1)$ conditions should be tested in order to judge whether a
state is separable or not. Though in mathematical form this method is more
complex than the one provided in \cite{7}, it is practical in judging the
separability of a state. Meanwhile, our theorem provides a method to find
the disentangled decomposition of a separable state. It is also worth
mentioning that for $2\times 2$ systems, the condition that the partial
transposition(PT) is positive and the condition that each $a^1\leq 0$ in our
method are equivalent. This suggests that there may be an essential
connection between the PT of $\rho $ and $a^1.$

\acknowledgments  We thank William K. Wootters and Guangcan Guo for their
encouragement. This work was supported in part by the National Natural Science
Foundation of China.

\begin{center}
APPENDIX
\end{center}

Considering an $m\times n$ system, a state $\rho $ of this system, which has 
$l$ nonzero eigenvalues, has a decomposition, $\rho =\sum_i\left|
y_i\right\rangle \left\langle y_i\right| ,$ which satisfies $\left\langle
y_i\right| \left. \widetilde{y}_j^r\right\rangle =\lambda _i\delta _{ij},$
where $\left| \widetilde{y}_i^r\right\rangle =B^r\left| y_i^{*}\right\rangle 
$ and $B^r$ is a transformation matrix corresponding to a parallel pair ($%
r=1,2,\cdots ,(m-1)(n-1)$). If $\lambda _1\leq \sum_{i=2}^{l^{\prime
}}\lambda _i$ ($l^{\prime }$ is the number of nonzero eigenvalues of $\tau
^r(\tau ^r)^{*}$), one can find a decomposition of $\rho ,$%
\begin{equation}
\left| z_i\right\rangle =\frac 1{2\sqrt{k}}\sum_{j=1}^la_{ij}e^{i\theta
_j}\left| y_j\right\rangle ,\quad i=1,2,\cdots ,4k,\quad k=1,2,3,\cdots
\label{bon}
\end{equation}
where $4k$ is greater than or equal to $l,$ $\left| a_{ij}\right| =1$ and
their signs are decided by Fig. 1, $\left| y_{l^{\prime }+1}\right\rangle
,\left| y_{l^{\prime }+2}\right\rangle ,\cdots ,\left| y_l\right\rangle $
being zero vectors. Phase factors can be found to make $\sum_je^{2i\theta
_j}\lambda _j=0$\cite{9}, which guarantees $\left\langle z_i\right| \left. 
\widetilde{z}_i^r\right\rangle =0.$ Obviously, the matrices $\left[
a_{ij}\right] $ is a $4k\times l$ matrix.

From Fig. 1, one can find that 
\[
a_{4\times 4}=\left[ 
\begin{array}{llll}
+1 & +1 & +1 & +1 \\ 
+1 & +1 & -1 & -1 \\ 
+1 & -1 & +1 & -1 \\ 
+1 & -1 & -1 & +1
\end{array}
\right] . 
\]
\begin{equation}
a_{8\times 8}=\left[ 
\begin{tabular}{llllllll}
$+1$ & $+1$ & $+1$ & $+1$ & $+1$ & $+1$ & $+1$ & $+1$ \\ 
$+1$ & $+1$ & $+1$ & $+1$ & $-1$ & $-1$ & $-1$ & $-1$ \\ 
$+1$ & $+1$ & $-1$ & $-1$ & $+1$ & $+1$ & $-1$ & $-1$ \\ 
$+1$ & $+1$ & $-1$ & $-1$ & $-1$ & $-1$ & $+1$ & $+1$ \\ 
$+1$ & $-1$ & $+1$ & $-1$ & $+1$ & $-1$ & $+1$ & $-1$ \\ 
$+1$ & $-1$ & $+1$ & $-1$ & $-1$ & $+1$ & $-1$ & $+1$ \\ 
$+1$ & $-1$ & $-1$ & $+1$ & $+1$ & $-1$ & $-1$ & $+1$ \\ 
$+1$ & $-1$ & $-1$ & $+1$ & $-1$ & $+1$ & $+1$ & $-1$%
\end{tabular}
\right] .  \label{918}
\end{equation}
According to the symmetry, one can find the higher-dimension matrix $%
a_{4k\times 4k}$ easily. If the number $l$ of nonzero eigenvalues of $\rho $
is less than $4k$, one may take the first $l$ columns of $a_{4k\times 4k}.$
For example, 
\[
a_{8\times 5}=\left[ 
\begin{tabular}{lllll}
$+1$ & $+1$ & $+1$ & $+1$ & $+1$ \\ 
$+1$ & $+1$ & $+1$ & $+1$ & $-1$ \\ 
$+1$ & $+1$ & $-1$ & $-1$ & $+1$ \\ 
$+1$ & $+1$ & $-1$ & $-1$ & $-1$ \\ 
$+1$ & $-1$ & $+1$ & $-1$ & $+1$ \\ 
$+1$ & $-1$ & $+1$ & $-1$ & $-1$ \\ 
$+1$ & $-1$ & $-1$ & $+1$ & $+1$ \\ 
$+1$ & $-1$ & $-1$ & $+1$ & $-1$%
\end{tabular}
\right] . 
\]

\newpage {\bf Figure Captions} \vspace{2ex}

\begin{center}
\begin{minipage}{120mm}
{\sf Fig. 1.} \small{The diagram showing the sign of the 
oefficient $a_{kj}$ in Eq. (16).}
\end{minipage}
\begin{figure}[htbp]
\begin{center}
\setlength{\unitlength}{1truecm}
\begin{picture}(6.8,6.8)
\put(-8.0,-15)
{\includegraphics{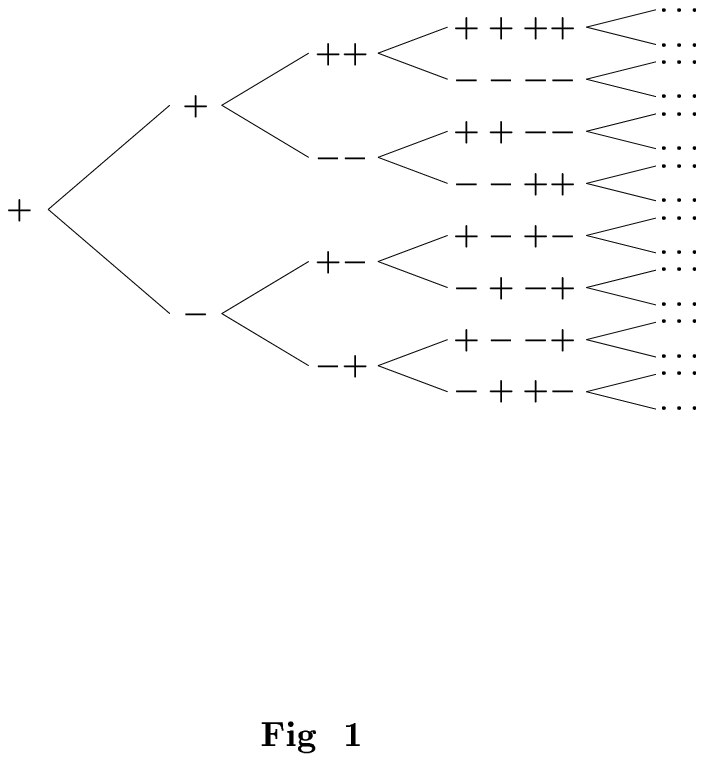}}
\end{picture}
\end{center}
\vskip 2.0cm
\protect\label{Fig.1}
\end{figure}
\end{center}

\end{document}